\documentclass{article}
\usepackage{arxiv}

\usepackage[utf8]{inputenc} 
\usepackage[T1]{fontenc}    
\usepackage{hyperref}       
\usepackage{url}            
\usepackage{booktabs}       
\usepackage{amsfonts}       
\usepackage{nicefrac}       
\usepackage{microtype}      
\usepackage{lipsum}
\usepackage{graphicx}
\graphicspath{ {./images/} }
\usepackage[linesnumbered,ruled,vlined]{algorithm2e}
\usepackage{amsmath,amsfonts}
\usepackage{algpseudocode}
\usepackage{graphicx}
\usepackage{textcomp}
\usepackage{xcolor}
\def\BibTeX{{\rm B\kern-.05em{\sc i\kern-.025em b}\kern-.08em
		T\kern-.1667em\lower.7ex\hbox{E}\kern-.125emX}}
\usepackage{pifont}
\usepackage{enumitem}
\usepackage{hyperref}
\usepackage[linesnumbered,ruled,vlined]{algorithm2e}

\usepackage{multirow}
\usepackage{booktabs}
\usepackage{dcolumn}
\usepackage{rotating} 
\usepackage{arydshln}
\usepackage{listings}
\SetKwInput{KwInput}{Input}                
\SetKwInput{KwOutput}{Output}
\usepackage{balance}
\usepackage{url}
\usepackage{hhline}
\usepackage{multirow}
\usepackage[most]{tcolorbox}

\tcbset{
    frame code={}
    center title,
    left=4pt,
    right=4pt,
    top=4pt,
    bottom=4pt,
    colback=gray!15,
    colframe=white,
    width=\dimexpr\textwidth\relax,
    enlarge left by=0mm,
    boxsep=5pt,
    arc=0pt,outer arc=0pt,
}
\urldef{\mailsa}\path||    
\usepackage{color}
\usepackage{tikz}
\usepackage{empheq}
\usepackage{mdframed}

\usetikzlibrary{calc, shapes.geometric, arrows, positioning}
\definecolor{mygreen}{rgb}{0,0.6,0}
\definecolor{mygray}{rgb}{0.5,0.5,0.5}
\definecolor{mymauve}{rgb}{0.58,0,0.82}

\newcommand{\tool}{GUPPY}

\def\code#1{\texttt{#1}}

\title{Automated Update of Android Deprecated API Usages with Large Language Models}

\author{
 Tarek Mahmud \\
  Texas State University\\
  San Marcos, Texas, USA \\
  \texttt{tarek\_mahmud@txstate.edu} \\
   \And
 Bin Duan \\
  The University of Queensland \\ Brisbane, Australia \\
  \texttt{b.duan@uq.edu.au} \\
  \And
 Meiru Che \\
  Central Queensland University \\ Brisbane, Australia.\\
  \texttt{m.che@cqu.edu.au} \\
  \And
 Awatif Yasmin\\
  Texas State University\\
  San Marcos, Texas, USA \\
  \texttt{nuc4@txstate.edu} \\
  \And
 Anne H. H. Ngu \\
  Texas State University\\
  San Marcos, Texas, USA \\
  \texttt{angu@txstate.edu}\\
  \And
 Guowei Yang \\
  The University of Queensland \\
  Brisbane, Australia \\
  \texttt{guowei.yang@uq.edu.au} \\
}

\begin{document}
\maketitle
\begin{abstract}
Android apps rely on application programming interfaces (APIs) to
access various functionalities of Android devices.
These APIs however are regularly updated to incorporate new features while the old APIs get deprecated. Even though the importance of updating
deprecated API usages with the recommended replacement APIs has been widely recognized, it is non-trivial to update the
deprecated API usages. Therefore,  the usages of deprecated APIs linger in Android apps and cause compatibility issues in practice.
This paper introduces \tool, an automated approach that utilizes large language models (LLMs) to update Android deprecated API usages. By employing
carefully crafted prompts, \tool\ leverages GPT-4, one of the
most powerful LLMs, to update deprecated-API usages, ensuring
compatibility in both the old and new API levels. Additionally, \tool\
uses GPT-4 to generate tests, identify incorrect updates, and refine
the API usage through an iterative process until the tests pass or a
specified limit is reached. Our evaluation, conducted on 360
benchmark API usages from 20 deprecated APIs and an additional 156
deprecated API usages from the latest API levels 33 and 34,
demonstrates \tool's advantages over the state-of-the-art techniques.
\end{abstract}
	
\section{Introduction} \label{introduction}
The mobile technology realm has evolved to become an indispensable facet of our daily lives. Mobile devices cater to diverse functionalities, from navigation and social interaction on platforms like social media to educational apps and immersive gaming experiences. Among the vast array of mobile operating systems, Android distinctly dominates, holding a 
market share of over 70\%\cite{market-share}. 
Android's extensive adoption has led to the Google Play Store 
hosting over 3.5 million Android apps as of 2022\cite{app}. These apps are developed using the Android application programming interface (API). Along with the Android operating system, these APIs get updated to let the developers use new and updated features in their apps. With each iteration of the Android API levels, new API methods and fields are introduced while existing ones get changed or removed. As of now, the latest version of the Android API level is 34. 

However, this API evolution brings challenges. In particular, Android API deprecation, where APIs are marked as deprecated and are often replaced with more efficient replacement APIs — introduces layers of complexities for developers. The usage of deprecated APIs can linger in apps, potentially leading to compatibility issues. The Android framework adheres to a "deprecate-replace-remove" cycle \cite{li2018characterising} that ensures that developers are equipped with the necessary resources and time to update these deprecated API usages to their recommended alternatives or replacement APIs. A grace period is typically extended before these deprecated APIs are eventually removed from the framework. This structured approach underscores the importance of updating the deprecated API usages, enabling developers to align with the API updates to enhance the quality of their apps.
Recent studies have identified the challenges associated with API deprecation and proposed automatic update of deprecated API usages in the app code \cite{fazzini2019automated, haryono2020automatic, haryono2021androevolve, thung2019towards, thung2020automated}. For example, Haryono et al. \cite{haryono2021androevolve} proposed an approach to updating deprecated API usages, but the approach requires proper update examples and cannot update all types of API usages. 
Zhao. et al. \cite{zhao2022towards} proposed a template-based method to automatically detect and fix compatibility issues in Android apps, but the method requires manual effort to generate those templates. 

{In this paper, we propose \tool, an approach to updating deprecated API usages with large language models (LLMs) for Android apps.  
In particular, \tool\ leverages GPT-4, one of the most powerful LLMs, to update the deprecated API usages and make them compatible with both old and new Android API levels. Furthermore, \tool\ also leverages GPT-4 to generate Robolectric~\cite{robolectric} tests to quickly refute incorrectly updated API usages. The generated tests are executed on both the old and new API levels,
and if the tests fail on either API level, \tool\ uses the failure information to refine the API usages or generate new tests. This iterative process continues until the tests pass or a user-specified bound is reached. We evaluate the performance of \tool\ by comparing it against the state-of-the-art technique AndroEvolve in updating two sets of deprecated API usages: 360 benchmark API usages of 20 deprecated APIs from AndroEvolve's study and a new set of 156 deprecated API usages from 156 deprecated APIs we collected from the two latest API levels 33 and 34. The experimental results show that \tool\ correctly updated 331 out of 360 benchmark deprecated API usages and 139 out of 156 more recent deprecated API usages, while AndroEvolve correctly updated only 301 and 83 deprecated API usages respectively.

This paper makes the following contributions:
\begin{itemize}
\item \textbf{Approach.} \tool, an approach to updating deprecated API usages for Android apps. 
To the best of our knowledge, \tool\ is the first approach that leverages LLMs for updating deprecated API usages. 
\item \textbf{Evaluation.}  An experimental evaluation of {\tool}, which shows that {\tool} outperforms the state-of-the-art technique AndroEvolve in updating benchmark deprecated API usages and more recent deprecated API usages.
\item \textbf{Data.} A public release of the new set with 156 deprecated APIs from the two latest Android API levels 33 and 34, source code~\cite{replication} of this work to facilitate the replication of our study and its application in more extensive contexts. 
\end{itemize}

\section{Background} \label{background}

\subsection{Large Language Models}
Large Language Models (LLMs) \cite{brown2020language} have become popular nowadays due to improvements in Natural Language Processing (NLP) \cite{nlp} that allow these models to be trained on lots of data. LLMs are designed with extensive general knowledge, and are either fine-tuned or given specific prompts to perform specific tasks. Fine-tuning adjusts the models with more training to perform a specific task, but can be expensive and sometimes lacks enough data to do it. On the other hand, prompting is simpler, and it provides the model with clear instructions
and maybe a few examples and then let the model perform the task.

LLMs are based on the transformer architecture \cite{singh2021nlp}. There are different kinds of LLMs. Some LLMs, like Codex \cite{codex}, predict the next word based on previous words. Others, like CodeBERT \cite{feng2020codebert}, are trained to fill in missing words based on the surrounding context. There are also models trained using reinforcement learning to teach models based on feedback. Examples are InstructGPT \cite{instructGPT} and ChatGPT \cite{chatGPT}, which start with a basic model and then adjust the model based on feedback from humans. They use a process called Reinforcement Learning from Human Feedback (RLHF) \cite{rlhf}, where they first adjust the model with some examples and then train a separate model to score the results, which further helps improve the main model. ChatGPT, in particular, is known for good conversations because it is trained on conversations and remembers past chats.

\subsection{Updating Deprecated API Usages}
Deprecated APIs~\cite{deprecation} are outdated interfaces that are discouraged to use because they can lead to compatibility and security issues. They are common in software development, including Android app development~\cite{deprecation}. 
As the Android API evolves to a new level, developers must update their apps by replacing deprecated APIs with the recommended replacement APIs.
Neglecting these updates can result in long-term problems, as deprecated APIs may eventually be removed, causing apps to have compatibility issues. When a deprecated API is replaced, the replacement API is often designed to be more efficient, more secure, or more compatible with new versions of Android. Developers are encouraged to use replacement APIs instead of deprecated ones to ensure that their apps remain up-to-date and compatible with future Android versions.

Listing \ref{lstExample} shows an example of deprecated and replacement API. In the Android API level $21$, the method \code{getExternalStorageDirectory()}, which gets the path to the primary shared/external storage directory, was marked as deprecated. Developers are recommended to switch to the \code{getExternalFilesDir()} method, i.e., the replacement API, which provides a more secure and flexible way to access the external storage directory.

\begin{lstlisting}[caption={An example of deprecated and recommended API to get the path of the external storage.}, label=lstExample]
// Deprecated API to get external storage directory
File externalStorageDir = Environment.getExternalStorageDirectory();
	
// Recommended API to get external storage directory
File externalFilesDir = getExternalFilesDir(null);
\end{lstlisting}

If developers continue to use the deprecated method \code{getExternalStorageDirectory()}, the app may still function correctly on devices running Android API $21$. However, if the app is run on devices running a new version of Android that no longer supports the deprecated method, the app may crash or fail to function correctly. To avoid this issue, developers should update their code to use the replacement API \code{getExternalFilesDir()}, which is supported on all Android API levels from 21 and above. Listing \ref{lstExampleUpdate} shows how to use the deprecated and replacement API in the same app.

\begin{lstlisting}[caption={How the deprecated API usages should be updated in the app code.}, label=lstExampleUpdate]
if(Build.VERSION.SDK_INT <= 21) {
	// Deprecated API usage
	File externalStorageDir = Environment.getExternalStorageDirectory();
}
else {
	// Recommended API usage
	File externalFilesDir = getExternalFilesDir(null);
}
\end{lstlisting}

Therefore, it is important to regularly check for deprecated APIs in Android apps and update them as necessary with recommended replacement APIs to prevent issues from deprecated APIs and to ensure optimal user experience and performance.

\subsection{Robolectric}
Robolectric~\cite{robolectric} is a fast and reliable unit testing framework for Android.  
Unlike traditional emulator-based methods, Robolectric runs tests in a specialized sandbox environment, allowing for fine-tuned Android configurations and ensuring tests do not interfere with each other. This framework provides enhanced test APIs for intricate control over Android behaviors. However, certain Android components, like hardware sensors, do not translate easily to this setting. Robolectric addresses this by offering test doubles, which, though not exact, effectively simulate these components for a majority of unit testing scenarios.

Listing \ref{lstRobolectricExampleCode} shows an example of an Android API that interacts with the Toast class to display a short message to the user. Testing this code is not straightforward with traditional unit testing frameworks, like JUnit as it involves UI rendering, but it is a scenario where Robolectric can be beneficial.

\begin{lstlisting}[caption={An API usage that is challenging to be tested with JUnit.}, label=lstRobolectricExampleCode]
public class ToastUtil {
    public static void showToast(Context context, String message) {
        Toast.makeText(context, message, Toast.LENGTH_SHORT).show();
    }
}
\end{lstlisting}

Listing \ref{lstRobolectricExampleTest} shows an example Robolectric test, where the ShadowToast class provided by Robolectric allows us to inspect the properties of the Toast, like the message it is supposed to display. This way, we can test API-related functionality that would otherwise be challenging with traditional unit tests.

\begin{lstlisting}[caption={A Robolectric test to check the API usage in Listing 3.}, label=lstRobolectricExampleTest]
@RunWith(RobolectricTestRunner.class)
@Config(manifest = Config.NONE)
public class ToastTest {
    private Activity activity;

    @Before
    public void setUp() {
        activity = Robolectric.buildActivity(Activity.class)
                                            .create().get();
    }

    @Test
    public void shouldDisplayToastWithCorrectMessage() {
        // Given
        String expectedMessage = "Hello Robolectric!";
        // When
        ToastUtil.showToast(activity, expectedMessage);
        // Then
        ShadowToast shadowToast = ShadowToast.getLatestToast();
        assertNotNull("Toast was not shown", shadowToast);
        assertEquals(expectedMessage, ShadowToast
                                .getTextOfLatestToast());
    }
}
\end{lstlisting}

\section{\tool} \label{approach}

Fig. \ref{fig:overview} shows the overview of \tool. 
\tool\ uses GPT-4 to update the deprecated Android API usage first. Then a test is generated by prompting GPT-4. The updated API usage is integrated into the app code and then tested on both old and new API levels. If the test fails on either API level, GPT-4 is prompted with the failure information to refine the API usage or generate a new test. \tool\ continues this process iteratively until the test passes on both Android API levels or a user-specified bound is reached.

\tool\ takes as input the deprecated API usage, the deprecated API signature, the API level in which it was deprecated, and the recommended replacement API signature. It consists of three main components: 1) Deprecated API Usage Update, 2) Test Generation, and 3) Iterative Refinement.

\subsection{Deprecated API Usage Update} \label{updateGeneration}
\subsubsection{Prompt Engineering}
{We first formulate a prompt for GPT-4, aiming to update the deprecated API usages to their recommended replacement APIs. 
The performance of this component lies in the crafting of a prompt tailored specifically for GPT-4. 
A key requirement throughout this update is to ensure that the updated deprecated API usage provided by GPT-4 remains functional across the diverse range of Android API levels.

Our prompt structure is designed to be both informative and directive. The template of the prompt is as follows:

\begin{tcolorbox} [width=470pt]
\small{Update the usage of Deprecated API \textbf{[Deprecated API]} in the following code. It is deprecated in API level \textbf{[API level]} and replaced with \textbf{[Replacement API]}. The updated code should be designed to maintain compatibility with both old and new Android versions. Please provide the updated code segment.

\textbf{[Code snippet]}}
\end{tcolorbox}

\begin{figure}[t!]
	\begin{center}
		\includegraphics[width=0.75\textwidth]{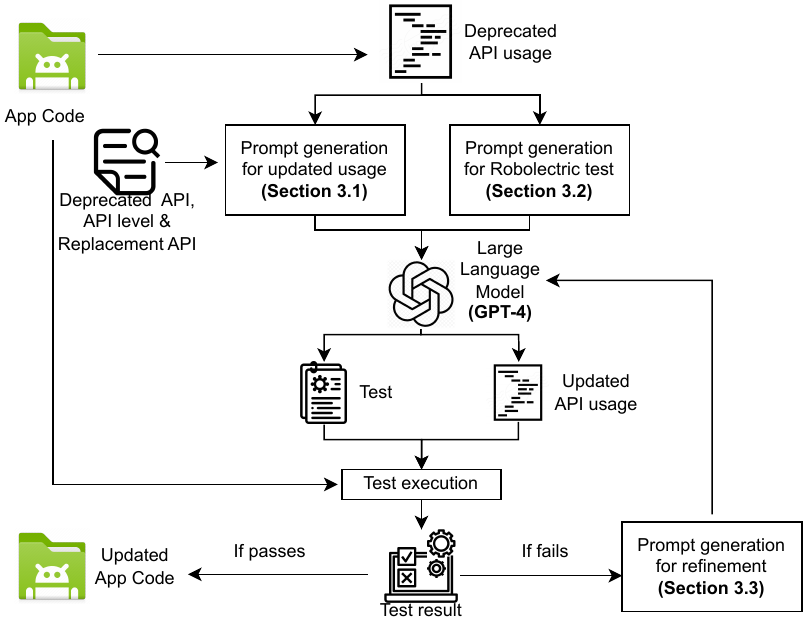}
		\caption{The overview of \tool.}
		\label{fig:overview}
	\end{center}
\end{figure}

To contextualize the prompt for specific instances, placeholders such as [Code snippet], [Deprecated API], [API level] and [Replacement API] are dynamically replaced with the actual code segment exhibiting the deprecated API usage, the exact signature of the deprecated API, the API level in which the deprecated API were marked as deprecated and the signature of its recommended replacement API, respectively.}

GPT-4 works best when being given clear details. How well it answers depends on the information it is provided. It uses this information to produce responses based on what it has learned. By specifying the deprecated API, its replacement, API level, and the code snippet, we are providing GPT-4 with a comprehensive view of our requirements and sharpening GPT-4's focus, so that it has necessary details to generate a solution. This eliminates potential ambiguities and guides the model towards our desired outcome. Furthermore, mentioning the need for compatibility across Android API levels is crucial. Without this directive, GPT-4 might provide a solution optimized solely for newer API levels, inadvertently neglecting older ones. Given that GPT-4 operates based on patterns and information it has been trained on, a clear and explicit prompt ensures that the solution generated aligns with the developer's intentions and the broader needs of the Android development ecosystem.

Moreover, the explicit nature of the prompt leverages GPT-4's vast training data and its understanding of best coding practices, ensuring the generated update does not just replace the deprecated API, but does so in a manner that is in line with industry standards. As shown in Listing \ref{lstPromtExample1}, when updating deprecated API usages in Android, developers usually keep both the deprecated and replacement API usages in the same code to make the update compatible across old and new Android API levels.

For instance, the prompt to update from the deprecated TimePicker.getCurrentHour() to its recommended replacement API TimePicker.getHour() in an example usage is as follows:

\begin{tcolorbox} [width=470pt]
    \small{Update the usage of Deprecated API \textbf{TimePicker.getCurrentHour()} in the following code. It is deprecated in \textbf{API level 23} and replaced with \textbf{TimePicker.getHour()}. The updated code should be designed to maintain compatibility with both old and new Android versions. Please provide the updated code segment.
    
    \textbf{int hour = timePicker.getCurrentHour()}}
\end{tcolorbox}

The generated output of GPT-4 for this prompt is shown in Listing \ref{lstPromtExample1}. This is compatible with both the old and new Android API levels.

\begin{lstlisting}[caption={Example deprecated API usage update with comprehensive prompt.}, label=lstPromtExample1]
if (Build.VERSION.SDK_INT >= Build.VERSION_CODES.M) {
    // Preferred method for Android 6.0 (Marshmallow/API level 23)
    // and above.
    int currentHour = timepicker.getHour(); 
} else {
    // Legacy method for versions below Android 6.0.
    int currentHour = timepicker.getCurrentHour(); 
}
\end{lstlisting}

On the other hand, when tried with fewer directive prompts like removing the API level and not mentioning the need for compatibility across Android API levels, GPT-4 generated ambiguous results. For the following prompt, 
\begin{tcolorbox} [width=470pt]
    \small{Update the usage of Deprecated API \textbf{TimePicker.getCurrentHour()} in the following code. It is replaced with \textbf{TimePicker.getHour()}. Please provide the updated code segment.
    
    \textbf{int hour = timePicker.getCurrentHour()}}
\end{tcolorbox}
the output generated by GPT-4 is shown in Listing \ref{lstPromtExample2}, which was not compatible with the old Android API level where TimePicker,getHour() was not available.

\begin{lstlisting}[caption={Example deprecated API usage update with simple prompt.}, label=lstPromtExample2]
// Introduced in Android 6.0 (Marshmallow/API level 23).
int currentHour = timepicker.getHour(); 
\end{lstlisting}


When there is a need to use multiple replacement APIs, we provided them separated by “and”. For example, ConnectivityManager.getAllNetworkInfo() API was deprecated in API level 23 and the recommended replacement APIs are ConnectivityManager.getAllNetworks() and ConnectivityManager.getNetworkInfo(android.net.Network). In this case, the prompt will be:

\begin{tcolorbox} [width=470pt]
    \small{Update the usage of Deprecated API \textbf{ConnectivityManager.getAllNetworkInfo()} in the following code. It is deprecated in API level \textbf{23} and replaced with \textbf{ConnectivityManager.getAllNetworks() and ConnectivityManager.getNetworkInfo(android.net.Network)}. The updated code should be designed to maintain compatibility with both old and new Android versions. Please provide the updated code segment.
    
    \textbf{[Code snippet]}}
\end{tcolorbox}

\subsubsection{API Usage Update}
The crafted prompt is then dispatched to GPT-4, setting the stage for the model's computational prowess. This is not just a matter of direct substitution. Thanks to the prompt, the model is acutely aware that the newly generated update needs to serve two purposes: directly replacing the deprecated API and ensuring compatibility across Android API levels. After obtaining the updated API usage from GPT-4, it is placed into the app code.

GPT-4 might provide different results for the same prompt and we did not set the temperature to control this type of situation. We attempted GPT-4 multiple times with the same input and found that GPT-4 generated correct updates most of the time even if the generated code is different. Even if the generated result is incorrect, the iterative refinement (described in Section \ref{iterative}) of our approach can mitigate that effect.

\subsection{Test Generation}
This component leverages GPT-4 to create a Robolectric~\cite{robolectric} test. By sending a tailored prompt that encapsulates the necessary details of the function and its context, \tool\ directs the model to generate a test that can evaluate the behavior of the inputted deprecated API usage. Note that only one test is generated as it is intended to quickly identify the wrongly updated code generated by GPT-4, rather than thoroughly testing the updated API usage. This test sets a clear baseline, by which we can find the difference between how the function originally worked and how it performs after the update.

This work uses Robolectric for its ability to facilitate testing across various Android API levels, which is important for our work, given that deprecated APIs often span multiple Android API levels, and their replacements might exhibit diverse behaviors across these API levels.

The Robolectric test is expected to capture the behavior of a function so that we can use it as a unit test, {thus} 
we pass to the prompt the name of the function that has the deprecated API usage. The prompt is carefully designed to convey this information in the following template:

\begin{tcolorbox} [width=470pt]
\small{Generate a Robolectric test for the function \textbf{[Function name]} to test its current functionality where the deprecated API \textbf{[Deprecated API]} is used. This will help us verify any changes made to its behavior after the update.

\textbf{[Code snippet]}}
\end{tcolorbox}

In this prompt, placeholders such as [Function name], [Deprecated API], and [Code snippet] are replaced with the specific function under consideration, deprecated API, and the input deprecated API usage respectively. This ensures that GPT-4 is equipped with the precise context it needs. Then we pass the prompt to GPT-4. This is a pretty straightforward process. The response from GPT-4 for the provided prompt is in the form of a detailed Robolectric test.  

\begin{lstlisting}[caption={Example Robolectric test to test similar code in different API levels.}, label=lstRobolectricAPILevelsExample]
@RunWith(RobolectricTestRunner.class)
public class UpdatedFunctionTest {

    @Test
    @Config(sdk = Build.VERSION_CODES.Q) // For API level 29
    public void testFunctionForAPI29() {
        // Test logic for API 29
    }

    @Test
    @Config(sdk = Build.VERSION_CODES.R) // For API level 30
    public void testFunctionForAPI30() {
        // Test logic for API 30
    }
}
\end{lstlisting}

After that, we integrate the updated deprecated API usage and the Robolectric test generated by GPT-4 into the app code. With Robolectric tests, a developer can effortlessly specify the desired API level for a particular test using the @Config annotation. For instance, if we intend to test the update generated by GPT-4 against both Android API levels 29 and 30, we could structure the test as shown in {Listing \ref{lstRobolectricAPILevelsExample} by running the same test for two different API levels}.

We employ such a structure by putting the test generated by \tool\ with two different configs to test if the updated API usage works as expected at two API levels before and after the API deprecation. {Testing for only two API levels, especially one API level before and one API level after API deprecation can confirm the API usage's compatibility across Android API levels \cite{xia2020android}. After integrating the test and updating API usage in the app, \tool\ runs the test. If the test successfully passes on both Android API levels, it indicates that the updated API usage is compatible with both the old and new API levels.}

\subsection{Iterative Refinement} \label{iterative}
This component refines the usage based on the test outcome.  For a failed test, we generate a prompt using test failure information to refine the previously generated deprecated API usage update. If the test fails on the old Android API level, we take the name of the test that failed and the error message and generate a new prompt for GPT-4. The template of this prompt is:

\begin{tcolorbox} [width=470pt]
\small{The Robolectric test named \textbf{[Test Name]} failed on older Android versions with the following error: \textbf{[Error Message]}. Please refine the test to correct this error.

\textbf{[Test code snippet]}}
\end{tcolorbox}

\tool\ tries again by giving GPT-4 this information, refining the test to work correctly. 
 
On the other hand, if the test passes for the old API level, but not on the new one, \tool\ takes the name and error message of the test to create a different prompt for GPT-4. The test can fail for the new API level for two possible reasons: (1) the updated API usage provided by GPT-4 is not correct and requires changes; (2) the replacement API may behave differently than the deprecated API. We create a prompt so that, if the replacement API has a different behavior that cannot be tested with the same test as the deprecated API usage, GPT-4 will come up with a new test to capture this behavior. Otherwise, GPT-4 will make changes in the updated API usage. The template of this prompt is:

\begin{tcolorbox} [width=470pt]
\small{The Robolectric test named \textbf{[Test Name]} passed on older Android versions but failed on newer ones with the error: \textbf{[Error Message]}. If the new API \textbf{[Replacement API]} behaves differently than the deprecated one \textbf{[Deprecated API]} and cannot be tested with the initially generated test, generate a new Robolectric test to capture this behavior. Otherwise, make changes in the code you provided to resolve the issue.

\textbf{[Test code snippet]}}
\end{tcolorbox}

After getting the updated usage or test from GPT-4, we integrate it into our app again. Then, we run the test on both the old and new Android API levels again. We iteratively provide test failure information to perform the refinement until the test passes on both old and new Android API levels or a user-specified bound is reached. If the test fails for any API levels after the specified bound of iterations is reached, we consider the update unsuccessful; otherwise, we consider the update successful.

	\newcommand{\rqOne}{RQ1: How does \tool\ perform in updating Android deprecated API usages against the state-of-the-art?}

\newcommand{\rqTwo}{RQ2: How effective is the chosen prompt in guiding \tool\ to update deprecated API usage?}

\newcommand{\rqThree}{RQ3: How do the test generation and iterative refinement of \tool\ perform?}

\section{Evaluation} \label{evaluation}

\subsection{Research Questions}
Our study aims to investigate the effectiveness of \tool\ against the state-of-the-art tools in updating Android deprecated API usages. To anchor our investigation, we pose the following three research questions:
\begin{itemize}[leftmargin=*]
\item \textit{\rqOne} 
\item \textit{\rqTwo} 
\item \textit{\rqThree} 
\end{itemize}

\subsection{Implementation}
We prompt GPT-4 to produce the updated API usage, where we provide it with the entire class file as the code snippet which includes all the codes and comments. We execute the primary logic of our approach using Python scripts, specifically designed to craft prompts based on the input parameters. Access to GPT-4 is achieved through its API version "gpt-4-0613". Once GPT-4 generates the updated API usage, we automatically rewrite the file in the app code. For the tests, we create a new test file in the test folder, and modify the test generated by GPT-4 as shown in Listing \ref{lstRobolectricAPILevelsExample} and run it for both the old and new API levels. Only the installation of plugins of Robolectric and Mockito for the Android Projects are performed manually, as different projects with different gradle versions may require different versions of these plugins. Updating all the projects to a single gradle version seems time-consuming as that may cause issues with other third-party library dependencies. We leave the automation of this part for our future work.

Throughout this evaluation, we maintain the authenticity of GPT-4's outputs, ensuring that our evaluation is a genuine reflection of GPT-4's capabilities. For the iterative refinement, we allow up to 5 iterations after generating the first test for the evaluation purpose. If the test continues to fail after the fifth iteration, it is concluded that \tool\ fails to update the deprecated API usage.

All the experiments of this study were conducted on a PC running Windows 10 64-bit, with an i5 Quad-Core 2.50 GHz CPU and 8 GB of RAM. We used the following tools to build the Android apps: Java SE Development Kit 11, Android Software Development Kit 32, and Android Studio Flamingo (2022.2.1).

\subsection{Baselines and Subjects}

We position the effectiveness of our approach in a comparative analysis with AndroEvolve, the state-of-the-art technique in the field. 
{Several tools have been proposed to update deprecated API usages for Android apps. 
AppEvolve~\cite{fazzini2019automated} is the first work that learns from code examples to update the deprecated API usage update. Like AppEvolve, A3~\cite{lamothe2020a3} also requires the target code under repair written syntactically similar to the before- and after-update API change examples. This makes these two hard to find applicable updates, as claimed by Thung et al. \cite{thung2020automated}. To address their limitations Haryono et al.~\cite{haryono2020automatic} proposed CocciEvolve, which is designed as an enhancement to AppEvolve. CocciEvolve outperformed AppEvolve on 112 targeted deprecated API usages. Building on this success, the same authors introduced AndroEvolve, refining and extending the capabilities of CocciEvolve, leading to better results in updating deprecated API usages. Thus, our comparative study focuses on comparing \tool\ with AndroEvolve.}

{In terms of the subjects used for the evaluation, we first use AndroEvolve's benchmark, which includes a set of 360 usages of 20 deprecated APIs. We meticulously sourced the officially recommended replacement API for each of these 20 deprecated APIs from Android developers' documentation \cite{androidDevelopers}. Moreover, we try beyond this benchmark and evaluate \tool's competence with a set of more recently deprecated APIs. The rationale behind this extended exploration stems from the understanding that GPT-4, having been trained until September 2021, might already be familiar with the deprecated API updates presented in the AndroEvolve benchmark. Thus, we ventured into the realm of APIs deprecated in the two latest API levels 33 and 34.}

From the comprehensive Android API differences reports for levels 33 (\url{https://developer.android.com/sdk/api_diff/33/changes}) and 34 (\url{https://developer.android.com/sdk/api_diff/34/changes}), we curated a list of 156 deprecated APIs. To apply AndroEvolve, we have also collected the replacement API, a usage and an update example for each of those deprecated APIs. Similar to Haryono et al. \cite{haryono2021androevolve}, we also selected public GitHub projects for the API usages and the update examples using AUSearch~\cite{asyrofi2020ausearch}.

\subsection{Results and Analysis}
\begin{table*}[t!]
\caption{Comparative results on AndroEvolve benchmark.}
\label{tab:rq1}
\centering
\scalebox{.7}{
\begin{tabular}{|l|c|c|c|}
\hline
API                                                                                                                & \#Usages & \tool\ & AndroEvolve \\ \hline
android.app.Notification.Builder\#addAction(int, java.lang.CharSequence, android.app.PendingIntent)                & 2        & 2                                    & 0           \\ \hline
android.content.ContentProviderClient\#release()                                                                   & 11       & 11                                   & 11          \\ \hline
android.graphics.Canvas\#saveLayer(float, float, float, float, android.graphics.Paint, int)                        & 21       & 17                                   & 10          \\ \hline
android.location.LocationManager\#addGpsStatusListener(android.location.GpsStatus.Listener)                        & 10       & 0                                    & 0           \\ \hline
android.location.LocationManager\#removeGpsStatusListener(android.location.GpsStatus.Listener)                     & 5        & 0                                    & 0           \\ \hline
android.media.AudioManager\#abandonAudioFocus(android.media.AudioManager.OnAudioFocusChangeListener)               & 1        & 1                                    & 0           \\ \hline
android.media.AudioManager\#requestAudioFocus(android.media.AudioManager.OnAudioFocusChangeListener, int, int)     & 53       & 48                                   & 32          \\ \hline
android.media.MediaPlayer\#setAudioStreamType(int)                                                                 & 2        & 2                                    & 0           \\ \hline
android.net.ConnectivityManager\#getAllNetworkInfo()                                                               & 11       & 9                                    & 4           \\ \hline
android.os.Vibrator\#vibrate(long)                                                                                 & 8        & 8                                    & 8           \\ \hline
android.os.Vibrator\#vibrate(long\{{[}\}\{{]}\}, int)                                                              & 6        & 6                                    & 6           \\ \hline
android.telephony.TelephonyManager\#getDeviceId()                                                                  & 29       & 27                                   & 29          \\ \hline
android.text.Html\#fromHtml(java.lang.String)                                                                      & 15       & 15                                   & 15          \\ \hline
android.view.View\#startDrag(android.content.ClipData, android.view.View.DragShadowBuilder, java.lang.Object, int) & 4        & 3                                    & 4           \\ \hline
android.webkit.WebViewClient\#shouldOverrideUrlLoading(android.webkit.WebView, java.lang.String)                   & 0        & 0                                    & 0           \\ \hline
android.widget.TextView\#setTextAppearance(android.content.Context, int)                                           & 15       & 15                                   & 15          \\ \hline
android.widget.TimePicker\#getCurrentHour()                                                                        & 60       & 60                                   & 60          \\ \hline
android.widget.TimePicker\#getCurrentMinute()                                                                      & 60       & 60                                   & 60          \\ \hline
android.widget.TimePicker\#setCurrentHour(java.lang.Integer)                                                       & 32       & 32                                   & 32          \\ \hline
android.widget.TimePicker\#setCurrentMinute(java.lang.Integer)                                                     & 15       & 15                                   & 15          \\ \hline
Total                                                                                                              & 360      & 331                                  & 301         \\ \hline
\end{tabular}}
\end{table*}

\subsubsection{\rqOne} 

To answer RQ1, we first applied \tool\ and AndroEvolve to update all the 360 deprecated API usages in the AndroEvolve benchmark. If the test passes, we manually analyze the results to check if there is anything wrong with the code or test, and then build and run the app with the updated API usage to draw a conclusion. This manual validation involves two critical checks. Firstly, we examine whether the replacement API is used within the updated code. If so, we proceed with further checking on test cases. If not, we mark it as a wrongly generated update. Secondly, we focus on the test cases associated with the API. If the replacement API usage is not available in the updated code, we mark it as a wrongly generated update. It is essential to verify that the tests effectively call the functions where both deprecated and replacement APIs are employed. This verification ensures that the tests are not only passing but are also calling the functions in a manner that is representative of actual usage scenarios, thereby validating the efficacy of the update. Similarly, we check the updates generated by AndroEvolve.

Table \ref{tab:rq1} shows the results of the updates generated by \tool\ against AndroEvolve in the AndroEvolve benchmark. Overall, AndroEvolve correctly updated 301 of the 360 deprecated APIs, while \tool\ demonstrated its effectiveness by correctly updating a total of 331. While both tools faced challenges in handling certain deprecated APIs, \tool\ shows broader adaptability. Although the tests passed for all 346 cases, when manually checked all these API updates, we found 15 usages of two APIs: \url{android.location.LocationManager\#addGpsStatusListener(android.location.GpsStatus.Listener)} and \url{android.location.LocationManager\#removeGpsStatusListener(android.location.GpsStatus.Listener)}, where \tool\ generates incomplete updates. However, the generated comments in the code can confirm the incompleteness of these two updates. AndroEvolve, also could not manage to generate an accurate fix for these methods.

\begin{lstlisting}[caption={Updated code example generated by \tool.}, label=lstAddGpsStatusListener]
public void addGpsStatusListener(GpsStatus.Listener listener) {
    mGpsStatusListener = listener;
    if (Build.VERSION.SDK_INT >= Build.VERSION_CODES.N) {
        // New method using GnssStatus.Callback
        mGnssStatusCallback = new GnssStatus.Callback() {
            @Override
            public void onSatelliteStatusChanged(GnssStatus status) {
              // You can call the appropriate methods on your 
              // original listener here.
              // This is just a basic example.
              mGpsStatusListener.onGpsStatusChanged(
                            GpsStatus.GPS_EVENT_SATELLITE_STATUS);
            }

            // You can also override other callback methods if needed
        };
        mLocationManager.registerGnssStatusCallback(
                                        mGnssStatusCallback);
    } else {
        // Old method for compatibility
        mLocationManager.addGpsStatusListener(mGpsStatusListener);
    }
}
\end{lstlisting}

The primary reason behind this challenge is the shift from the listener approach used in these methods to a newer callback mechanism. This change mandates the creation of a fresh object for the callback. In its attempt to address this, \tool\ crafted a solution where it initialized a callback object and then added a comment. This comment was strategically placed to guide the developer, suggesting that they integrate whatever actions the original listener object was designed to perform. An illustration of \tool's approach for the method \url{android.location.LocationManager\#addGpsStatusListener(android.location.GpsStatus.Listener)} can be seen in Listing \ref{lstAddGpsStatusListener}.

In contrast, AndroEvolve's solution was less comprehensive. For the said method, it merely declared a callback variable but omitted its initialization. As a result, although this variable was then passed to the \url{android.location.LocationManager\#registerGnssStatusCallback(GnssStatus.Callback)} method and the code compiled successfully, it presents a lurking issue. Listing \ref{lstAddGpsStatusListener1} shows the updated API usage generated by AndroEvolve. Specifically, the incomplete nature of this update means that when this code is executed in a real-world scenario, it would inevitably result in a NullPointerException, making the app unstable.

\begin{lstlisting}[caption={Updated code example generated by AndroEvolve.}, label=lstAddGpsStatusListener1]
private GnssStatus.Callback callback;
public void addGpsStatusListener(GpsStatus.Listener listener) {
    mGpsStatusListener = listener;
    if (android.os.Build.VERSION.SDK_INT >= 
                        android.os.Build.VERSION_CODES.N) {
        locationManager.registerGnssStatusCallback(callback);
    }
    else{
        locationManager.addGpsStatusListener(mGpsStatusListener);
    }
}
\end{lstlisting}

\begin{lstlisting}[caption={Test generated by \tool\ to test the deprecated API usage.}, label=lstAddGpsStatusListenerTest]
@RunWith(RobolectricTestRunner.class)
public class MixedPositionProviderTest {
    private MixedPositionProvider mMixedPositionProvider;
    private LocationManager mLocationManager;
    private ShadowLocationManager mShadowLocationManager;

    @Before
    public void setUp() {
        mMixedPositionProvider = new MixedPositionProvider();
        mLocationManager = mock(LocationManager.class); 
        mShadowLocationManager = Shadow.extract(mLocationManager);
    }

    @Test
    public void testGpsStatusListener() {
        GpsStatus.Listener listener = mock(GpsStatus.Listener.class);
        mMixedPositionProvider.addGpsStatusListener(listener);
        verify(mLocationManager).addGpsStatusListener(listener);
        verifyNoMoreInteractions(mLocationManager);
    }
}
\end{lstlisting}

Listing \ref{lstAddGpsStatusListenerTest} presents a test example created by \tool, designed to validate the deprecated API usage of the updated usage in Listings \ref{lstAddGpsStatusListener} and \ref{lstAddGpsStatusListener1}. This cannot test the updated version as the GpsStatus.Listener is not registered in the updated API usage.

When asked to generate different tests or refine the updated API usage, \tool\ generates two tests shown in Listing \ref{lstAddGpsStatusListenerTest2}. One for versions less than Android N (using the deprecated API), which is similar to the previous test, and another for versions greater or equal to Android N. This test suite validates the addGpsStatusListener method for different Android versions. The testGpsStatusListenerBeforeN checks its function for versions before Nougat, ensuring the older GpsStatus.Listener is used. On the other hand, testGpsStatusListenerAfterN verifies the use of the modern GnssStatus.Callback for Nougat and later versions.

\begin{lstlisting}[caption={Test generated by \tool\ to test the updated code.}, label=lstAddGpsStatusListenerTest2]
@RunWith(RobolectricTestRunner.class)
public class MixedPositionProviderTest {
    private MixedPositionProvider mMixedPositionProvider;
    private LocationManager mLocationManager;
    private ShadowLocationManager mShadowLocationManager;

    @Before
    public void setUp() {
        mMixedPositionProvider = new MixedPositionProvider();
        mLocationManager = mock(LocationManager.class); 
        mShadowLocationManager = Shadow.extract(mLocationManager);
    }

    @Test
    @Config(sdk = Build.VERSION_CODES.M) // Before Android N
    public void testGpsStatusListenerBeforeN() {
        GpsStatus.Listener listener = mock(GpsStatus.Listener.class);
        mMixedPositionProvider.addGpsStatusListener(listener);
        verify(mLocationManager).addGpsStatusListener(listener);
        verifyNoMoreInteractions(mLocationManager);
    }

    @Test
    @Config(sdk = Build.VERSION_CODES.O) // After Android N
    public void testGpsStatusListenerAfterN() {
        GpsStatus.Listener listener = mock(GpsStatus.Listener.class);
        mMixedPositionProvider.addGpsStatusListener(listener);

        // Here, we could further verify the behavior of 
        // mGnssStatusCallback by simulating a callback
        verify(mLocationManager).registerGnssStatusCallback(
                            any(GnssStatus.Callback.class));
        verifyNoMoreInteractions(mLocationManager);
    }
}
\end{lstlisting}

This test encountered a failure when applied to the updated API usage from Listing \ref{lstAddGpsStatusListener1}, which was produced by AndroEvolve. Conversely, it successfully passed for the API usages in Listings \ref{lstAddGpsStatusListener} and \ref{lstAddGpsStatusListener1}. It is worth noting, however, that the test ideally should have detected a flaw and failed for Listing \ref{lstAddGpsStatusListener} due to the inadequately implemented callback object within that updated API usage. As the test only checked if the GnssStatus.Callback is created or not, the test passes on this incomplete updated API usage.

When analyzing the failed cases generated by \tool,  especially for the APIs android.telephony.TelephonyManager\#getDeviceId and android.view.View\#startDrag, we identified two scenarios. First one is, GPT-4 generated incomplete code updates, even though the tests passed. If GPT-4 needs to generate a new object to replicate the attributes of another object that is not available within the same file, similar to the example shown above, then \tool\ generates an incomplete update. GPT-4 also failed when the code files contained multiple instances of the deprecated API. GPT-4 struggles to identify the correct context or sequence for API replacement when multiple similar instances occur within a single class. We have discussed these scenarios in detail in Section \ref{discussion}.

After that, we applied \tool\ and AndroEvolve to update the deprecated API usage of our new dataset collected from API levels 33 and 34. This presented a fresh set of complexities, as these APIs are relatively new and could potentially be outside the familiarity domain of both tools. 

From the total of 156 deprecated APIs in these levels, we find the absence of suitable update examples for 44 of these APIs. It indicates the limitation of AndroEvolve's capabilities as it is unable to cater to these 44 instances, given its reliance on pre-existing update examples. From the remaining 112 APIs, AndroEvolve managed to update 83 instances correctly.

On the other hand, \tool\ reported 151 cases where the tests passed. When analyzing the 5 failed cases, we found that the deprecated API usage code has multiple API usages of the mentioned API in the same inputted API Usage code snippet and \tool\ either failed to generate tests for all cases or generate updated codes for all cases in 5 iterations. 

When manually analyzing the updated code, we found 12 cases where \tool\ generates incomplete deprecated API usages like the example shown above. This consistent behavior across both datasets underscores a specific area of potential enhancement for \tool.

We were able to update these 516 deprecated API usages (from both datasets) within 108.45 hours with \tool. On average, the initial usage update takes 9.5 seconds, test generation takes 8.1 seconds, and each iteration of the refinement takes 7.6 seconds. The rest of the time was taken by the process of adding the updated usages and tests into the app code, building the Android projects, and running tests for each iteration. The usage of the OpenAI API costs us on average \$0.31 for each API update.

The experimental results show that \tool\ is more effective in updating a wider range of deprecated API usages than AndroEvolve.

\subsubsection{\rqTwo} \label{sec:rq2}

{To evaluate the effectiveness of our chosen prompt shown in Section \ref{updateGeneration}, we compared it with the following two alternative prompts with varying amounts of information in the prompt in a zero-shot \footnote{Zero-shot prompting refers to the model's ability to perform a task or generate responses without specific training or prior exposure to that task.}:

\begin{enumerate}
    \item \textbf{Prompt A:} This prompt is constructed with the deprecated API usage without any supplementary information. GPT-4 is tasked with autonomously recognizing and updating the deprecated usages. The prompt is as follows:

    \begin{tcolorbox} [width=435pt]
    \small{Update any deprecated Android API usages in the following code. The updated code should be designed to maintain compatibility with both old and new Android versions. Please provide the updated code segment.

    \textbf{[Code snippet]}}
    \end{tcolorbox}

    \item \textbf{Prompt B:} This prompt is constructed with the deprecated API's name alongside the deprecated API usage. This offers a clear directive, channeling GPT-4's focus on the specified API. It is as follows:
    \begin{tcolorbox} [width=435pt]
    \small{Update the usage of Deprecated API \textbf{[Deprecated API]} in the following code. The updated code should be designed to maintain compatibility with both old and new Android versions. Please provide the updated code segment.

    \textbf{[Code snippet]}}
    \end{tcolorbox}
\end{enumerate}}

{In this experiment, we passed the whole class with comments as a code snippet with the prompts to the GPT-4. After letting GPT-4 to update the deprecated API usage, we manually analyzed the changes (similar to answering RQ1) to ensure they were updated correctly for both datasets. 
As a result, we found that GPT-4 was able to update 273, 314, and 331 deprecated API usages in Prompt A, Prompt B, and our chosen prompt respectively for the AndroEvolve dataset.
On our new dataset, GPT-4 successfully updated 103, 124, and 139 deprecated API usages with Prompt A, Prompt B, and our chosen prompt respectively. These results underline the advantage of our chosen prompt in comparison with the alternative prompts.

In a detailed analysis of the instances where GPT-4 failed to generate correct updates with Prompt A, we identified that when tasked with simultaneously detecting and then updating deprecated API usages, GPT-4 sometimes struggled to manage these dual responsibilities effectively. This observation becomes particularly pronounced when we consider the length of the input files. 
Furthermore, a large number of 
code snippets provided to GPT-4 contained usages of multiple deprecated APIs, where  GPT-4 faces difficulty in recognizing and updating them all accurately within a single operation. On the other hand, Prompt A, Prompt B, and our chosen prompt all have difficulty in updating multiple usages of the target deprecated API in the code snippet.}


\subsubsection{\rqThree}

To answer RQ3, we checked how \tool\ performs in test generation and interactive refinement when quickly checking for issues in the generated updated code and refining them.
Here we checked how much of the initially generated tests were good and how it performs in iterative refinement of the tests and the updates.

When updating a total of 516 deprecated API usages from both datasets used in this evaluation, we found that 425 tests were successful. This achievement demonstrates that 82.36\% of the tests met the syntactical and structural criteria essential for a valid Robolectric test. A consistent issue faced by \tool\ in unit test generation was multiple usage of the provided API and the tests’ ability to handle updated code that was intertwined with varying objects or activity dependencies. However, \tool\ was able to generate tests within 5 iterations for all these cases that passed for the deprecated API usage in old Android API levels. The success rate speaks to \tool's capability to produce structurally sound tests, even when faced with the intricate requirements of Android API testing.

\begin{figure}[t!]
	\begin{center}
		\includegraphics[width=0.75\textwidth]{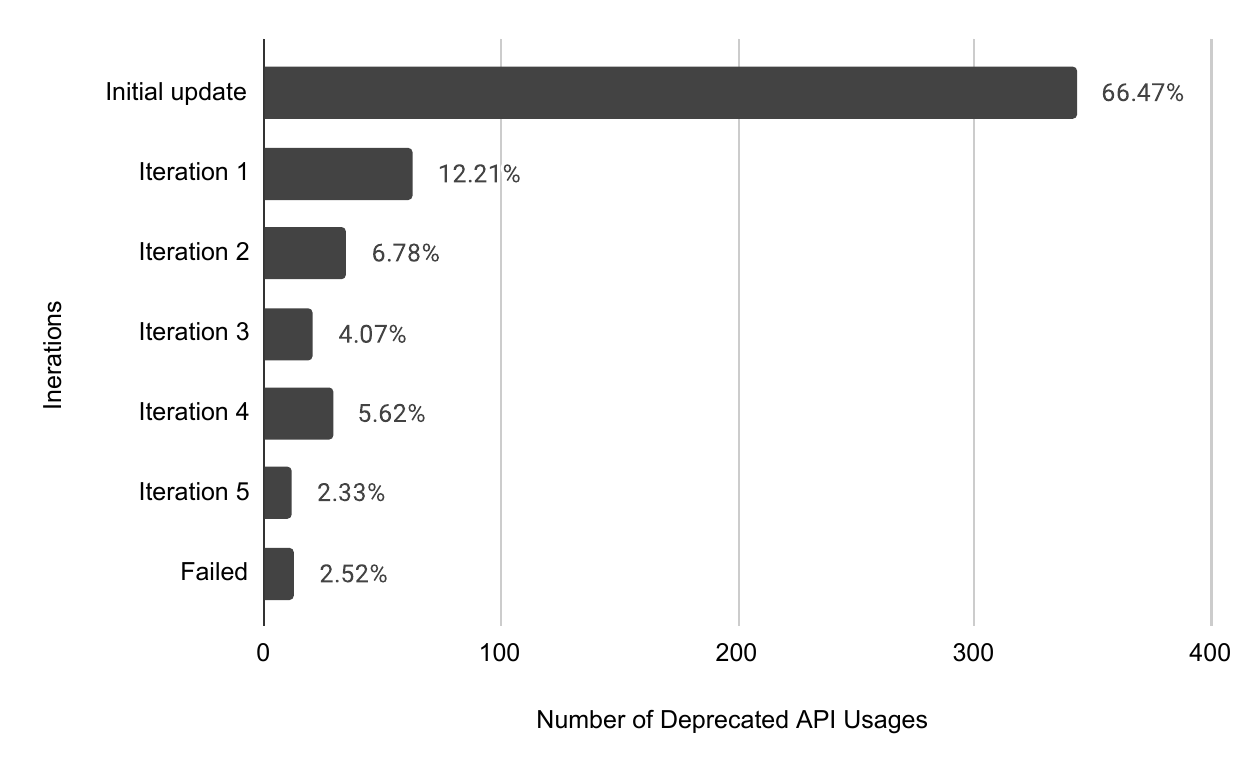}
		\caption{Iterations taken by \tool\ to update deprecated API usages.}
		\label{fig:iterations}
	\end{center}
\end{figure}

Fig. \ref{fig:iterations} shows the iterations needed to pass the test by \tool\ when updating these deprecated API usages. According to the figure, the baseline test was passed 343 updated usages out of 516 passed. In 64 other cases the initial deprecated API usage generated by \tool\ was correct, but the test either failed for faulty test generation or the replacement API had a different behavior. \tool\ was able to validate 63, 35, 21, 29, and 12 updated usages during the iteration 1 to 5 respectively. In only 13 cases the test did not pass in 5 iterations.

According to the results, even though the number of correctly updated deprecated API usage in zero-shot is high, the iterative testing-based refinement can refute many incorrectly updated usages and improve the performance of \tool.


        \section{Discussions} \label{discussion}

In this section, we investigated the 36 failed cases in total from both datasets during the evaluation of \tool\ and discussed the benefits \tool\ provides for using large language models. Here we try to correctly update them with the knowledge gathered from the comments of these failed updates. We noticed two scenarios of different scenarios of failure 23 incomplete updated codes where the tests passed and 13 failed API updates. 

We were able to determine the incompleteness of the update generated by \tool\ by reading the comments generated by GPT-4. The effectiveness of using a large language model is, that it not only provides the updated code for deprecated API usage, but it also provides comments that can be leveraged to fulfill the update. Here we tried to navigate and correctly update 23 cases where \tool\ generated incomplete updated code by following the comments. According to the comments of these 23 cases, the main challenge lay in situations where \tool\ had to generate a new object and mirror attributes of another object from the app code, especially if the referenced object was not available within the same file. When faced with such a challenge, we supplement GPT-4 with the exact generation code of the object that is needed to generate the correct updated usage. This directed input served as a bridge, guiding the GPT-4 to produce accurate updates by filling the contextual gap. Listing \ref{lstAddGpsStatusListenerUpdated} shows the example code of updating the \url{android.location.LocationManager\#addGpsStatusListener(android.location.GpsStatus.Listener)} method after providing the code of GpsStatus.Listener listener.

\begin{lstlisting}[caption={Updated code example generated by GPT-4.}, label=lstAddGpsStatusListenerUpdated]
public void addGpsStatusListener(GpsStatus.Listener listener) {
    mGpsStatusListener = listener;
    if (Build.VERSION.SDK_INT >= Build.VERSION_CODES.N) {
        // New method using GnssStatus.Callback
        mGnssStatusCallback = new GnssStatus.Callback() {
            @Override
            public void onSatelliteStatusChanged(GnssStatus status) {
                mKnownSatellites = status.getSatelliteCount();
                mUsedInLastFixSatellites = 0;
                for (int i = 0; i < mKnownSatellites; i++) {
                    if (status.usedInFix(i)) {
                        mUsedInLastFixSatellites++;
                    }
                }
            }
        };
        mLocationManager.registerGnssStatusCallback(mGnssStatusCallback);
    } else {
        // Old method for compatibility
        mLocationManager.addGpsStatusListener(mGpsStatusListener);
    }
}
\end{lstlisting}
 
Moreover, another set of challenges faced by \tool\ is the code files with multiple instances of the provided deprecated API. \tool\ failed 13 times in these scenarios, possibly struggling with finding the right context or sequence of API replacements when faced with multiple, similar instances within a single class. They have also mentioned the multiple usages in the comments. In such cases, when providing the entire class in the prompt, we mentioned the line number of the deprecated API usage. This isolation aimed to simplify the context, thereby aiding GPT-4 in making accurate API replacements without being overwhelmed by the multitude of similar instances.
        \section{Threats to Validity} \label{threats}
The primary threats to external validity pertain to the generalizability of our results to other LLMs. Our findings on GPT-4's performance might not generalize to other large language models or other versions of the GPT series. The challenges and nuances of Android API updates might not directly translate to other platforms or languages. Also, our study focuses on GPT-4, where LLMs' capabilities are rapidly evolving. Future versions or other models might perform differently. In this evaluation, we used the deprecated APIs from Android levels 33 and 34. Different results might be obtained if another dataset or a broader range of API levels are used.

Regarding internal validity, there is always a risk of bugs in our experimental setup, tool configuration, or Python scripts. We have tried our best to mitigate this by testing each module separately. {There is also a risk of GPT-4 hallucinating when generating the updated API usage as well as the test to validate that update. We manually checked the generated updates and tests, and did not find any such cases in our evaluation.}

Where threats to construct validity are concerned, we measured success in API usage updates, Robolectric test, and manual analysis. However, there are other potential metrics, like runtime performance or user experience, which we did not consider.
\section{Related Work} \label{relatedWork}
This section discusses the related works on API evolution and deprecation, and using large language models for software engineering.

\subsection{API Evolution and Deprecation}
Lamothe et al. \cite{lamothe2021systematic} conducted a systematic review of API evolution research, identifying six primary research areas that focus on API evolution. Linares et al. \cite{linares2013api, linares2014api} found that successful apps tend to use APIs that are less prone to faults and changes compared to unsuccessful apps and Android API changes lead to a higher number of Stack Overflow discussions.
McDonnell et al. \cite{mcdonnell2013empirical} found that API updates are more vulnerable than other changes. Mutchler et al. \cite{mutchler2016target} examined security risks in older Android apps on newer devices, while Li et al. \cite{li2016accessing} tracked inaccessible Android APIs by examining the code base. Yang et al. \cite{Yang:2018:AOS:3197231.3197258} analyzed the impact of Android API updates on apps.
Several techniques have been proposed to detect API evolution-induced compatibility issues \cite{li2018cid, huang2018understanding, he2018understanding, acid, xia2020android}.



In software systems, APIs are commonly deprecated to introduce new features or address issues, ideally following a deprecate-replace-remove cycle \cite{10.1145/2950290.2950298, 1510134, 10.5555/1133105.1133107, 10.1145/1932682.1869518}. However, Zhou et al. \cite{10.1145/2950290.2950298} observed that in open-source Java frameworks and libraries, this cycle is often neglected.
Sawant et al. \cite{8453124, 8529833} discovered issues with the current Java deprecation mechanism not meeting developer needs, while Raemaekers et al. \cite{6975655} found cases of deprecated APIs persisting in Java systems, causing codebase inconsistencies, particularly in Android. Brito et al. \cite{BRITO2018306, 7476657} emphasized the importance of clear replacement messages for faster adoption of new APIs, supported by Ko et al. \cite{7091210} who found well-documented replacements lead to more updates. 
Li et al. \cite{li2018characterising} analyzed Android code and open-source apps to understand the frequency and reasons for deprecated API usage, providing insights into their impact. 

There are several works on the automatic update of deprecated API usage in Android. Lamothe et al. \cite{lamothe2020a3} introduced A3, an automated tool for simplifying Android API migrations by learning from code examples and providing migration solutions. Fazzini et al. \cite{fazzini2019automated} introduced AppEvlolve, an approach that automatically updates deprecated APIs in Android apps by learning from before- and after-update examples. However, concerns have been raised about the generalizability of AppEvolve \cite{thung2020automated}, leading to the proposal of CocciEvolve \cite{haryono2020automatic} and AndroEvolve \cite{haryono2021androevolve}, which address these issues by using data flow analysis and variable denormalization for efficient API updating. Haryono et al. \cite{9609153} developed MLCatchUp, which utilizes natural language processing to suggest replacements for deprecated machine learning APIs in Python based on semantic similarity. In this paper, we proposed \tool\ which is the first tool that leverages LLM in updating deprecated API usages.

\subsection{Large Language Models}
{Large language models (LLMs) have been widely used in many language tasks. It also exhibits impressive performance in numerous coding-related tasks such as code completion \cite{deng2022fuzzing, huang2022prompt, jain2022jigsaw, li2022competition, xu2022systematic}, code synthesis \cite{liu2023your} and program repair \cite{xia2023automated}. Codex \cite{codex}, used by Github Copilot \cite{copilot}, showed promise by turning simple comments into code, potentially saving developers time, though there are still minor issues with it. Other large language models like BERT\cite{devlin2018bert, tian2023best}, GPT \cite{gpt4, chatGPT, xia2023keep}, and others \cite{le2023invalidator, paul2023automated} have also proven they can produce code that is both correct and relevant. Other studies are looking at how LLMs change the way we make software \cite{imai2022github, liu2023fill}, help developers work faster \cite{dakhel2023github, ziegler2022productivity}, and find security problems \cite{pearce2023examining}.

Among the large language models, GPT stands out a bit more because of its design which allows back-and-forth communication.
This interactive approach, along with detailed comparisons to other methods, highlights its efficiency and adaptability. This paper uses an iterative refinement approach to bring a fresh perspective in updating deprecated Android API usages using LLMs.}
\section{Conclusion} \label{conclusion}

In conclusion, this paper introduces \tool, a novel approach leveraging large language models like GPT-4 to efficiently update deprecated API usages in Android apps and to quickly identify and refine wrong updates with iterative testing and refinement. We curated a comprehensive dataset of 156 deprecated APIs from Android API levels 33 and 34 and evaluated \tool\ against existing techniques which consistently showcased superior performance. 
In the future, we plan to evaluate \tool\ on a broader range of deprecated API usage updates in Android apps.


	

	\balance
	\bibliographystyle{plain}
	\bibliography{Bibliography}

\end{document}